\documentclass[12pt]{iopart}
\usepackage[utf8]{inputenc}
\usepackage{amssymb}
\usepackage{amssymb}
\usepackage{graphicx}
\usepackage{url}
\usepackage{cite}
\usepackage{times}
\usepackage{bm}
\DeclareBoldMathCommand{\bfmu}{\mu}
\DeclareBoldMathCommand{\bfsigma}{\sigma}
\usepackage[svgnames]{xcolor}

\newcommand{\figpath}{}

\begin{document} 

\title{Stern-Gerlach splitting of low-energy ion beams}

\author{Carsten Henkel%
\footnote[7]{henkel@uni-potsdam.de}
}
\address{Institute of Physics and Astronomy, 
University of Potsdam, 
Karl-Liebknecht-Str. 24/25, 14476 Potsdam, Germany}

\author{Georg Jacob%
\footnote[8]{Present address: Alpine Quantum Technologies GmbH, Maria-Theresien-Stra{\ss}e 24, 6020 Innsbruck, Austria},
Felix Stopp, 
Ferdinand Schmidt-Kaler}
\address{QUANTUM, Institut f\"ur Physik,
Johannes-Gutenberg Universit\"at Mainz,
Staudingerweg 7, 55128 Mainz, Germany}

\author{Mark Keil, Yonathan Japha, Ron Folman}
\address{Department of Physics, 
Ben-Gurion University of the Negev, 
Beer-Sheva 84105, Israel}

\begin{abstract}
We present a feasibility study with several magnetic field configurations for creating spin-dependent forces that can 
 split a low-energy ion beam by the Stern-Gerlach effect. To the best of our knowledge, coherent spin-splittings of charged particles have yet to be realised. Our proposal is based on ion source parameters taken from a recent experiment that demonstrated single-ion implantation from a high-brightness ion source combined with a radio-frequency Paul trap. The inhomogeneous magnetic fields can be created by permanently magnetised microstructures or from current-carrying wires with sizes in the micron range, such as those recently used in a successful implementation of the Stern-Gerlach effect with neutral atoms. All relevant forces (Lorentz force and image charges) are taken into account, and measurable splittings are found by analytical and numerical calculations.
\end{abstract}


\submitto{\NJP}
\vfill\eject

\section*{Introduction}

The spin is a fundamental property of quantum particles, be they elementary or composite. First hints were provided by the discovery of the anomalous Zeeman effect, pre-dating even the Bohr atomic theory~\cite{Preston1898}: atomic electrons give an `anomalous' Zeeman shift because their spin magnetic moment
\begin{equation}
\bfmu = - g_e \mu_B \hat{\bf S}
\label{eq:g-factor}
\end{equation}
contains a Land\'e factor~$g_e\neq1$ which differs from the magnetic moment due to orbital angular momentum. Here~$\mu_B=e\hbar/2m_e$ is the Bohr magneton and~$\hat{\bf S}$ the (dimensionless) spin operator with eigenvalues~$S_z=\pm1/2$. Dirac's relativistic equation for the electron predicts~$g_e=2$, while corrections from quantum electrodynamics lead to~$g_e\approx2.00232\ldots$, in very good agreement with experiment. A direct experimental demonstration of the spin was the Stern-Gerlach~(SG) experiment~\cite{Gerlach1922} where a particle beam is split by a magnetic gradient, according to the spin-dependent force
\begin{equation}
{\bf F}_{\rm SG} = \nabla ( \bfmu \cdot {\bf B} )
= -\ \frac{ g_e e \hbar }{ 2 m_e } \sum_{i} S_i \nabla B_i
\,.
\label{eq:SG-force}
\end{equation}
The original experiment was performed~100 years ago by Otto Stern and Walther Gerlach with a beam of neutral atoms. The question whether this could also be done with charged particles like electrons was vigorously debated in the early days of quantum mechanics. Bohr and Pauli argued that spin splitting was impossible for a free electron beam on the basis of the~Uncertainty Relations. To see this, consider the Lorentz force
\numparts%
\begin{equation}
{\bf F}_L({\bf x}) = \frac{ e }{ m } {\bf p} \times {\bf B}({\bf x})
\,.
\label{eq:Lorentz-force}
\end{equation}
Take a beam of charged particles 
in an inhomogeneous magnetic field with momentum $p$ and spatial width $\Delta x$
along the magnetic gradient $B'$ (perpendicular to the beam axis).
The transverse component of the Lorentz force then broadens by
\begin{equation}
\Delta F_L = \frac{ e }{ m } p \Delta B = \frac{ e }{ m } p B' \Delta x
\,.
\end{equation}
By the uncertainty principle, one finds
\begin{equation}
\Delta F_L \ge \frac{e \hbar }{ 2 m } \frac{ p }{ \Delta p_x } B' 
> \frac{ m_e }{ m } \frac{e \hbar }{ 2 m_e } B' 
= \frac{ m_e }{ m } F_{\rm SG}
\,,
\label{eq:Lorentz-width}
\end{equation}\endnumparts%
where in the second inequality we have used that 
for a collimated beam, the transverse momentum width is obviously smaller
than the axial momentum, $p / \Delta p_x > 1$. 
For electrons ($m = m_e$, the electron mass),
the width in the Lorentz force
is therefore larger than the Stern-Gerlach 
splitting~[Eq.\,(\ref{eq:SG-force})].
For recent reviews of this issue, we refer to Batelaan~\cite{Batelaan2002} and Garraway and Stenholm~\cite{Garraway2002}. In contrast, if we take
ions such that $m_e / m < 10^{-3}$,
the lower limit given by Eq.\,(\ref{eq:Lorentz-width}) does not 
exclude~SG 
splitting, and this motivates our proposal for using a low-energy ion beam. The proposal is based on ground-state~$\rm^{40}Ca^+$ ions; with no nuclear spin and an alkali-like electronic configuration, its magnetic moment is dominated by the electron spin. Since ions are much more massive than electrons~($m\gg m_e$) and laser cooling can provide sub-mK temperatures, conditions can be found where the broadening due to the Lorentz force~Eq.\,(\ref{eq:Lorentz-force}) does not prevent spin-dependent splitting. 

In our feasability study, we assume the beam is generated by releasing ions from a miniaturised linear Paul trap. As characterised in recent experiments~\cite{Jacob2016}, the beam parameters have allowed resolving angular splittings of about~$\rm1\,mrad$. Our proposal utilises steep magnetic gradients, either from permanent magnets with sharp edges, or from patterned structures on a microchip that can be fabricated with state-of-the-art techniques~\cite{Armini2011,Keil2016}. In one of the latter configurations, the ions cross a magnetic grating where the direction of the field rotates along their trajectory. The~SG splitting then happens because the spin is ``wiggling'' in synchronisation with the field, similar to the proposal of Bloom and Erdman~\cite{Bloom1962}. In this example, the spin is far from being locked to a fixed quantisation axis, a situation that is quite typical because of strongly inhomogeneous fields.

The beam splitter for ion beams suggested here may form a basic building block of free space interferometric devices for charged particles. This would be 
similar to the electron interferometer of Hasselbach and co-workers~\cite{Sonnentag2007,Hasselbach2010} (which was not based on the spin degree of freedom), and in analogy to recently realised neutral particle~SG interferometers~\cite{Machluf2013,Margalit2018,Amit2019}. We anticipate that such a device could measure the coherence of spin splitting, putting ``Humpty-Dumpty together again'' (using the wording of~\cite{Englert1988,Schwinger1988,Scully1989}), and provide new insight concerning the fundamental question of whether and where in the~SG device a spin measurement takes place. The ion interference would also be sensitive to Aharanov-Bohm phase shifts arising from the electromagnetic gauge field. The ion source would be a truly single-particle device and eliminate certain problems arising from particle interactions in high-density sources of neutral bosons~\cite{TinoKasevich2014}. This opens a wide spectrum of fundamental experiments, probing for example weak measurements and Bohmian trajectories. The strong electric interactions may also be used, for example, to entangle the single ion with a solid-state quantum device (an electron in a quantum dot or on a Coulomb island, or a qubit flux gate). This type of interferometer may lead to new sensing capabilities~\cite{Hasselbach2010}: one of the two ion wave packets is expected to pass tens of microns above a surface (in the chip configuration of the proposal) and may probe van der Waals and Casimir-Polder forces, as well as patch potentials. These are very important as they are believed to give rise to the anomalous heating observed in miniaturised ion traps~\cite{Brownnutt2015}. Due to the short distances between the ions and the surface, the device may also be able to sense the gravitational force on small scales~\cite{Behunin2014}. Finally, such a single-ion interferometer may enable searches for exotic physics. These include spontaneous collapse models, the fifth force from a nearby surface, the self-charge interaction between the two ion wave packets, and so on. Eventually, one may be able to realise a double~SG splitter with different orientations, as originally attempted by Stern, Segr\`e, and co-workers~\cite{Phipps1932,Frisch1933}, in order to test ideas like the Bohm-Bub non-local hidden variable theory~\cite{Bohm1966,Das2019}, or ideas on deterministic quantum mechanics (see, e.g.,~\cite{Schulman2017}). Since ions may form the base of extremely accurate clocks, the~SG device suggested here would enable clock interferometry at a level sensitive to the Earth's gravitational redshift (see the proof-of-principle experiments with neutral atoms in Refs.~\cite{Margalit2015,Zhou2018}). This has important implications for studying the interface between quantum mechanics
and general relativity.


\section{Ion source}
\label{s:ion-source}

\begin{figure}[h]
\centerline{\includegraphics*[width = 0.7\textwidth]{\figpath 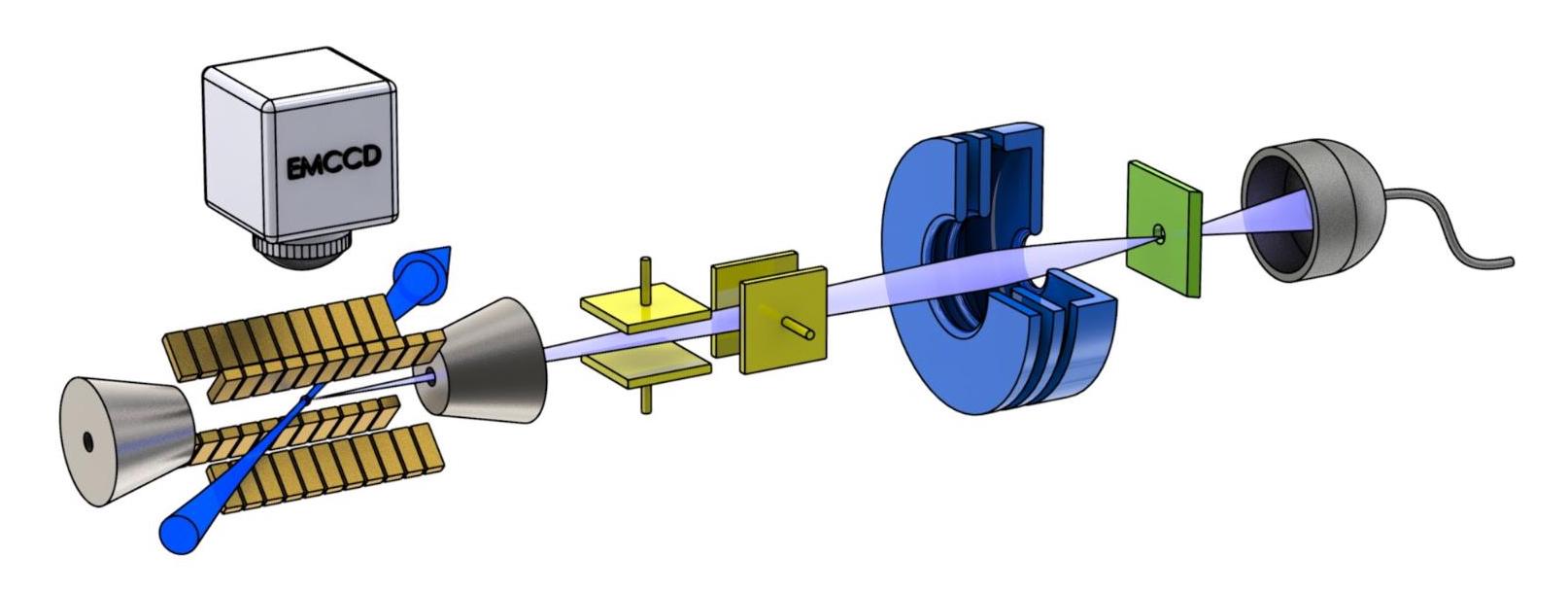}}
\caption[]{Sketch of the ion beam apparatus~\cite{Jacob2016}. From left: linear segmented ion trap closed by two (grey) endcaps with holes. Trapped ions are laser-cooled (blue arrow) and are detected by an~EMCCD camera. They are extracted through the right endcap by applying a voltage. The beam direction is adjusted with pairs of parallel plates (yellow), and focused by an electrostatic lens (blue). In the Stern-Gerlach~(SG) splitting experiment, the (green) pinhole would be replaced by a planar chip with micro-wires or by magnetised structures, typically oriented for grazing incidence.
}
\label{fig:source}
\end{figure}

\noindent
The ion beam apparatus is shown in Fig.~\,\ref{fig:source}. The ions are launched from a linear segmented Paul trap, employing radio-frequency~(RF) and~DC voltages~\cite{Meijer2006concept,schnitzler2009deterministic,izawa2010controlledExtraction}. The trap consists of four gold-coated alumina chips 
mounted in an X-shaped arrangement. The chips are segmented into~11 electrodes that shape the axial potential of the trap. The trap is first loaded with a number of $\rm^{40}Ca^+$ ions, produced by photoionisation from an atomic calcium beam. The trapped ions are Doppler cooled with laser light at~$\rm397\,nm$. After crystallising into a linear array, the number of ions is counted by means of an~EMCCD camera. The ions are then removed from the trap by changing 
its~DC potential until a single calcium ion remains. The single ion is launched through a dedicated hole in one of the endcaps of the trap by applying a voltage that may be chosen in the range of~$\rm0.3\ldots6\,keV$. Here we discuss the experimentally realised capabilities of the ion source and parameters that may be achievable with improvements, as well as more fundamental limits.

\begin{table}[h]
\caption[]{Characterisation of the ion source. The first column represents the current status, as used in Ref.~\cite{Jacob2016}. The values in the second column may be reached with improvements of the setup.}
\label{t:source}
\begin{indented}
\item[]%
\begin{tabular}{llll}
\br
                                       & current performance	                 & with modifications                    \\
\mr
extraction rate                        & $\rm3\,s^{-1}$                        & $\rm10^5\,s^{-1}$                     \\
lowest beam velocity $v$               & $\rm38\,\rm km/s$	                   & $\rm0.7\,km/s$                        \\
axial velocity spread $\delta v$	     & $\rm7.5\,m/s$	                       & $\rm0.7\,m/s$                         \\
angular divergence $\delta\theta$	     & $\rm23.7\pm2.5\,\mu rad$      	       & $\rm215\,\mu rad^{(a)}$ 	             \\
emittance (2D)$\rm^{(b)}$ 	           & $\rm2.6\,nm^2\,mrad^2\,eV$	           & $\rm0.13\,nm^2\,mrad^2\,eV$           \\
\br
\end{tabular}%
\item[$\rm^{(a)}$ ] The increased angular divergence is due to the slower beam. The estimate is for a trap temperature~$T\approx\rm44\,\mu K$ (0.24 phonons), achieved after sideband cooling in a~$\rm1.6\,MHz$ ion trap~\cite{Poschinger2009}.
\item[$\rm^{(b)}$ ] The~2D emittance is proportional to the product of the beam cross section and the transverse velocity spreads (transverse velocities and angles are related via the beam velocity) and is the product of both transverse~(1D) emittances. The inverse of the emittance is a measure of transverse phase-space density. A minimum uncertainty (or ``single mode'') beam has emittance~$\hbar^2/(8m)$, the value quoted in the second column. It is achievable by a fully adiabatic extraction from the trap ground state. 
\end{indented}
\end{table}

The source was characterised with respect to the extraction rate, the beam velocity and its spread, as well as its angular divergence and emittance (see Table~\ref{t:source} for the concept of the emittance and for typical values). The extraction rate of our source~($\rm3\,ions/s$) is currently limited by the ion loading rate. The largest value of the latter reported so far in ion traps was~$\rm4\cdot10^5\,s^{-1}$~\cite{Cetina2007}. In that work atoms were ionised from a magneto-optical trap superimposed with the Paul trap. Another limiting factor at higher extraction rates may be the detection of individual ions with the camera. The minimum required exposure time reported so far is~$\rm10\,\mu s$~\cite{Myerson2008}. This step could be made faster if the removal of excess ions was more reliable and control images were no longer needed. We note however that the extraction rate would affect only the experimental data-taking rate and not the instrument resolution.

The axial velocity~$v$ and its spread~$\delta v$ were determined by time-of-flight measurements. The lowest extraction voltage used in these experiments was~ $\rm300\,V$, corresponding to a~$\rm Ca^+$ velocity of~$v=\rm38\,km/s$ (earlier experiments~\cite{Schnitzler2009} used lower voltages but with much greater angular divergence). We anticipate that the greatest difficulty in reducing the ion energy may be fringe~RF fields of the ion trap. In simulations where the~RF is switched off immediately prior to applying the extraction voltage, we found that an energy as low as~$\rm0.1\,eV$ was observed, corresponding to~$v=\rm700\,m/s$. A new apparatus, designed for extraction even at low voltages, is currently being put into operation.

The velocity spread was found to be limited by noise in the high voltage switches used for applying the extraction voltage. Given that the lower-energy experiments do not require such high voltages, this could be improved by using low-voltage switches that provide lower noise characteristics. At a fundamental level, the velocity spread is limited by the motional energy uncertainty of the ion in the harmonic trap potential along the extraction axis. 

The beam angular divergence was experimentally determined by employing a profiling edge, which was successively stepped into the beam while recording the detector signal. From simulated ion trajectories, we found that one has to take into account the lensing effect of the electrical fields employed for the ion extraction. Consequently, reducing the anticipated extraction voltage to~$\rm0.1\,V$ also increases the angular divergence, due both to this additional lensing and to the larger ratio of transverse-to-axial velocity. The angular divergence is ultimately limited by the ground state kinetic energy transverse to the beam axis. 

The emittance of the beam (proportional to the product of the transverse widths in position and momentum) was inferred from spectroscopic temperature 
measurements performed on the~$\rm729\,nm$ quadrupole transition of~$\rm^{40}Ca^+$~\cite{schnitzler2009deterministic,izawa2010controlledExtraction}. As with the velocity spread, the emittance is ultimately limited by the ion's energy in the trap. For an ion cooled to its motional ground state, this leads to the value given in the second column of Table~\ref{t:source}.


\section{Magnetic gradient configurations}

The ion source has achieved unprecedented precision and control, and it has been applied to single-ion microscopy with~nm resolution~\cite{Jacob2016}. We now examine whether inhomogeneous magnetic fields, as used for Stern-Gerlach~(SG) experiments on neutral beams, can be used for spin-dependent manipulation of ion beams. We consider three different devices and their corresponding~SG splittings, which have to date been masked by the unavoidable presence of the large Lorentz broadening. We will also discuss the extent to which further improvements (second column of Table~\ref{t:source}) may be fruitful in achieving this goal. 


\subsection{Magnetised edges}
\label{s:mag-poles}

\begin{figure}[t]
\centerline{\includegraphics*[width = 0.8\textwidth]{\figpath 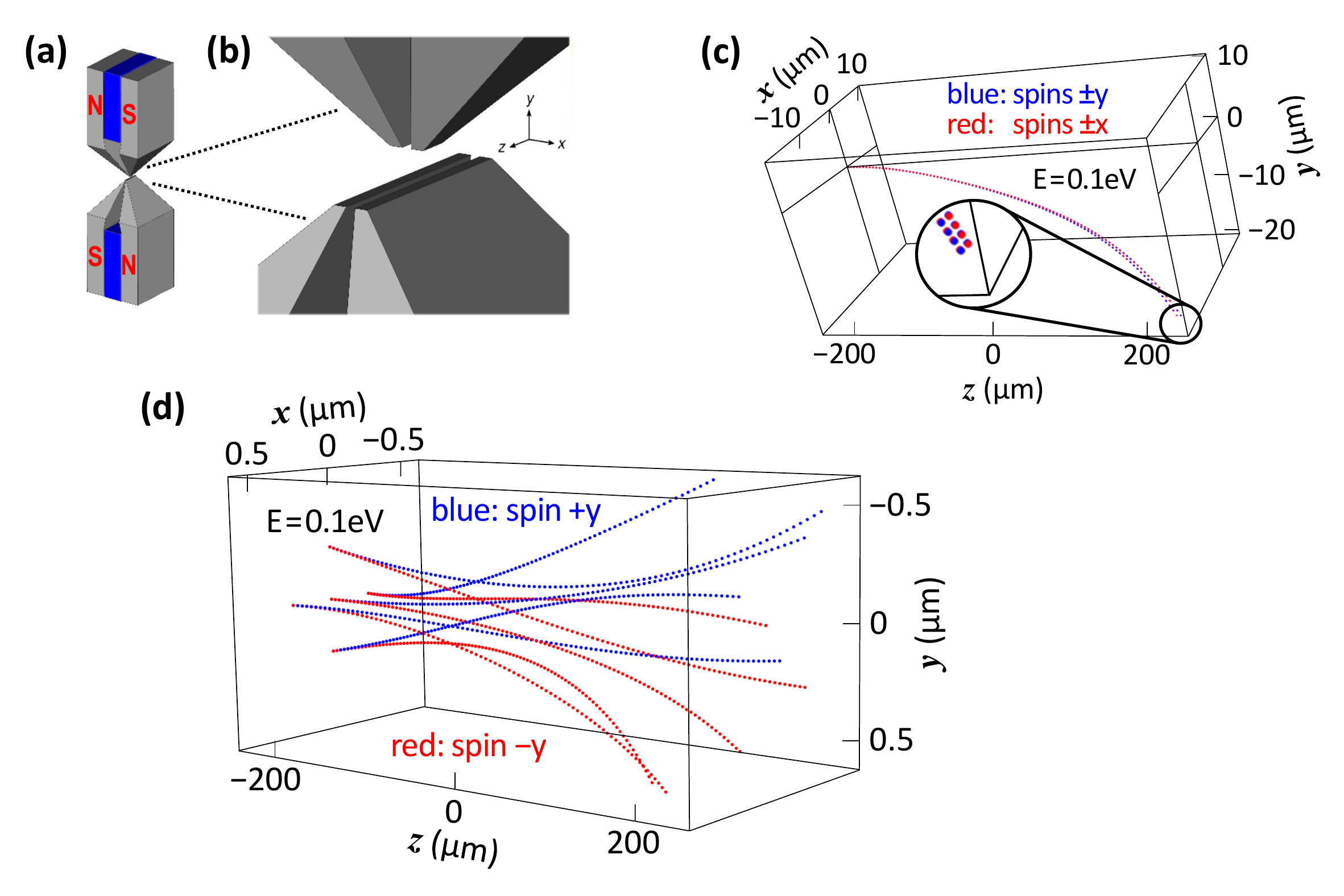}}
\caption[]{
(\emph{a,b})~Sharp magnetised pole pieces used in a gravito-magnetic trap for nano-diamonds~\cite{Hsu2016}. The pole pieces consist of~FeCo (grey) and~SmCo (blue) magnets. There is a~${\rm150\,\mu m}=2y_0$ spacing between the top and bottom poles, whose length along the~$z$\mbox{-}axis is in the~$\rm100\,\mu m$ range. The magnetic field is given by the potential of~Eq.\,(\ref{eq:multipole-exp}). The parameters~$a_2=\rm-1.3\,T$, $a_3=\rm-0.018\,T$, and~$a_4=\rm0.72\,T$ were found by fitting the magnetic potential to trapping frequencies observed for a trapped diamagnetic particle. The hexapole term~($a_3$) is a correction 
due to the top pole pieces being shorter than the bottom ones.
(\emph{c})~Trajectories of~$\rm^{40}Ca^+$ ions launched at~$\rm0.1\, eV$ along the~$z$-axis between the poles. Strong bending from the Lorentz force masks the spin-dependent splitting almost completely (inset).
(\emph{d})~Bundle of trajectories with initial conditions focusing them to the centre of the structure. The hexapole term~($a_3$) is removed by making the bottom and top magnetic poles symmetric. Deflection of off-axis trajectories by the Lorentz force is thereby reduced compared to~(\emph{c}) [note the~$40\times$ scale change along the~$x$ and~$y$ axes], while the spin-dependent splitting remains.
\\
Plots~(\emph{a,b}) have been adapted from Fig.~1 of Hsu et al., \emph{Scientific Reports}~{\bf6} (2016) 30125, Ref.~\cite{Hsu2016}, International License:
Creative Commons Attribution~4.0.
}
\label{fig:mag-wedge}
\end{figure}

Steep magnetic gradients can be created between magnetised pole pieces with sharp edges, as already used by Gerlach and Stern. The configuration sketched in~Fig.~\ref{fig:mag-wedge}(\emph{a, b}) was used in Ref.~\cite{Hsu2016} for generating very large gradients capable of trapping diamagnetic nano-diamonds. The static magnetic field outside the magnetised structures can be accurately computed from a scalar potential, ${\bf B}=-\nabla\Phi$. A multipole series for the latter can be written, up to an octupole term, as
\begin{equation}
\hspace*{-0.5\mathindent}
\Phi( {\bf r} ) = \frac{ a_2 }{ 2 y_0 }\
\sqrt{\frac{ 15 }{ 4\pi } }\ { x y }
+
\frac{ a_3 }{ 3 y_0^2 }\
\sqrt{\frac{ 21 }{ 32\pi } }\ { x(4z^2 - x^2 - y^2) }
+ \frac{ a_4 }{ 4 y_0^3 }\
\sqrt{\frac{ 315 }{ 16\pi } }\ { xy(x^2 - y^2) }
\,. 
\label{eq:multipole-exp}
\end{equation}
Here,~$a_2\ldots a_4$ (in~$\rm T$) specify the magnitudes of the quadrupole, hexapole, and octupole components, and~$y_0$ is a characteristic length (see figure caption for details and the choice of coordinates). Note that the field is zero at the origin.

The~SG force in this configuration has some unusual features, and we therefore provide a few technical details~\cite{next_theory_paper}. The most familiar expression for the spin-dependent force is probably the one in the adiabatic limit where the angle between the spin direction~${\bf S}$ and the magnetic field~${\bf B}$ remains fixed (e.g., the spin is parallel or anti-parallel to the field). In that case, Eq.(\ref{eq:SG-force}) becomes proportional to~$\nabla|{\bf B}|$, and the magnetic gradient gives the direction of the force.

Quite the opposite conditions apply to the configuration of Fig.\,\ref{fig:mag-wedge}, at least close to the~$z$-axis. By evaluating Eq.(\ref{eq:SG-force}) for the first term in the magnetic field (quadrupole, superscript~$2$), we get the force
\begin{equation}
{\bf F}^{(2)} = \mu B' 
\left( \begin{array}{c} 
	S_y \\ 
	S_x \\ 
	0 
\end{array} \right)
\label{eq:SG-force-quadrupole}
\end{equation}
where $B'\sim a_2/y_0\sim\rm10^4\,T/m$. A spin polarised along the positive~$y$-direction will be deflected by the~SG force along the~$x$-direction [see blue trajectories in Fig.\,\ref{fig:mag-wedge}(\emph{d})]. This deflection subjects the charged particle to the Lorentz force due to the nonzero magnetic field away from the~$z$ axis, thereby generating the relatively complex behaviour of the trajectories. In Fig.\,\ref{fig:mag-wedge}(\emph{c}), the hexapole term~[$\sim a_3$ in Eq.(\ref{eq:multipole-exp})] generates a nonzero magnetic field component~$B_x$ along the~$z$-axis (that vanishes only for~$z=0$), and the  bending due to the Lorentz force is then much larger than the spin-dependent splitting. Note that for both trajectories shown, the direction of the spin is kept constant. As already argued by Bloom and Erdman~\cite{Bloom1962}, this can be ensured by adding a homogeneous magnetic field (``bias field''). The spin component perpendicular to the bias then precesses and gives only an oscillating contribution to the force that does not generate a large deflection.

We note that in this analysis, we actually apply a semiclassical approximation and work with the expectation value of the spin operator (Ehrenfest theorem). The trajectory becomes deterministic and the approximation cannot describe the splitting of the ion beam. Nevertheless, the~SG force switches sign for the opposite spin orientation, as can be seen qualitatively in Fig.\,\ref{fig:mag-wedge}(\emph{d}). Another limitation of the model is that Eq.(\ref{eq:multipole-exp}) does not accurately describe the finite extent of the field that we expect to be concentrated in the region between the magnetic poles. If we artificially fix an interaction length~$L$, we may estimate the angular splitting using Eq.(\ref{eq:SG-force-quadrupole}) and the beam data from the second column of Table~\ref{t:source}:
\begin{equation}
\Delta \theta_x \simeq \frac{ \mu B' L }{ m v_z^2 }
\simeq 0.2\,{\rm mrad}\ \frac{ L }{ 100\,\mu{\rm m} }
\,,
\label{eq:mag-pole-splitting}
\end{equation}
which is consistent with the results shown in Fig.\,\ref{fig:mag-wedge}(\emph{d}). Compared to the beam divergence, we thus expect a well-resolved splitting from an idealised interaction region~$L$ of a few hundred microns.
 
As a simple estimation of the spin precession angle, let us consider an initial spin polarisation in the~$y$-direction and the magnetic field~$B_x$ existing on the symmetry axis~($x=y=0$) that arises from the hexapole:
\begin{equation}
B_x(0,0,z) = \frac{ a_3 }{ 3 }\
\sqrt{\frac{ 21 }{ 2\pi } }\ 
\frac{ z^2 }{ y_0^2 }
\,.
\label{eq:Bx-hexapole}
\end{equation}
Multiplying this by the Bohr magneton~$\mu_B/(2\pi\hbar)$, we get the frequency with which the spin rotates in the~$yz$-plane. Along the length~$L$, we get a total precession proportional to~$(a_3\mu_B/\hbar)L^3/(y_0^2 v)$ radians. For the parameters used here, this amounts to a few hundred rotations, thereby averaging out the~SG force due to the~$S_y$-component of the spin. We conclude that an experimental realisation using the configuration in Fig.\,\ref{fig:mag-wedge} would either implement a symmetric geometry to remove the hexapole term, or states whose spin is initially polarised along the~$\pm x$-axis. The second choice would clearly be inferior, as shown in Fig.\,\ref{fig:mag-wedge}(\emph{c}) where the splitting (in the~$y$-direction) is barely visible because it is superposed on the beam bending due to the Lorentz force. Broadening due to the spatially inhomogeneous Lorentz force will be estimated below.

The permanent-magnet configuration of this section is closest to the original~SG setup and has the advantage of a relatively simple design. Nevertheless, as noted earlier, our calculations neglect inaccuracies in the multipole expansion [Eq.(\ref{eq:multipole-exp})], even with the symmetric magnetic pole geometry, and we therefore consider two additional configurations in the following, whereby we benefit from modern chip fabrication techniques to design the magnetic field.


\subsection{Two wires}
\label{s:wire-pair}

\begin{figure}[t]
\centerline{%
\includegraphics*[width = 0.5\textwidth]{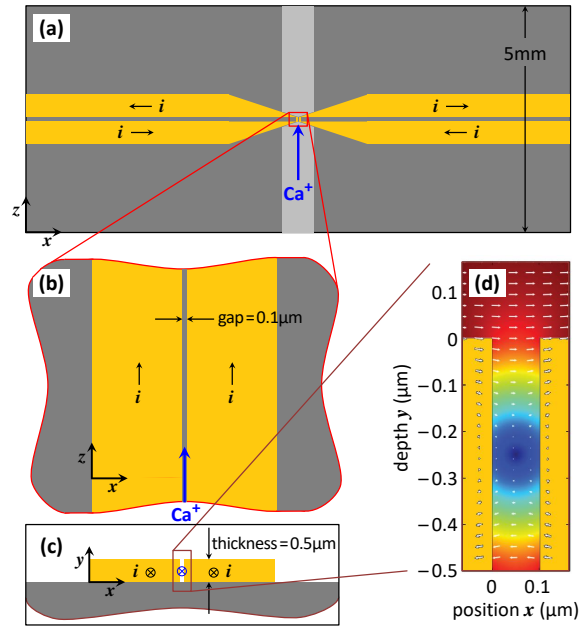}%
}
\caption[]{
(\emph{a})~Sketch of a microchip with a pair of wires parallel to the~$z$-axis (gold colour) deposited on a silicon substrate~(grey). The lighter shading corresponds to an area that must be milled down so that the~$\rm Ca^+$ ion avoids attractive surface forces for most of its trajectory. The direction of currents in the leads to each wire is shown.
(\emph{b-c})~Magnified views of the central region (to scale) comprising the parallel-wire configuration [top view in~(\emph{b}) and cross-section view from behind the ion beam in~(\emph{c})]. The~$\rm Ca^+$ ion beam would pass between the two gold wires, parallel to the~$z$-axis [blue mark in~(\emph{c})]. 
The~$y$-axis is normal to the chip surface; the top of the gold wires is at~$y=0$. 
(\emph{d})~Sketch of the magnetic field between the two wires (white arrows). The colours encode the magnitude of the field (blue is zero, red is large).
}
\label{fig:sketch-chips}
\end{figure}

Our two additional configurations are based on micron-sized wires fabricated on a planar substrate (so-called ion chips~\cite{Keil2016}). We begin with a very simple scheme that generates a magnetic gradient, namely a pair of parallel wires, each with current~$I$ and length~$L$, and whose centres are separated by a distance~$2d\ll L$ (see Fig.~\ref{fig:sketch-chips}). The magnetic field is then of the order of $\mu_0I/(2\pi d)$, where $\mu_0$ is the magnetic constant. For equal currents flowing in the same direction, the field, by design, is actually zero along the symmetry line between the wires [the dark blue region in Fig.~\ref{fig:sketch-chips}(\emph{d})], thus reducing the Lorentz force to a minimum. The magnetic gradient there is 
\begin{equation}
    B' = \frac{ \mu_0 I }{ \pi d^2 }
\,,
\label{eq:estimate-gradB}
\end{equation}
and can reach values up to a few~$\rm10^4\,T/m$ for realistic values listed in Table~\ref{t:chip-values}.

\begin{table}[t]\small
\begin{indented}
\item
\caption[]{Design values for generating inhomogeneous magnetic fields above a microchip. 
Wire pair (Sec.\,\ref{s:wire-pair}, Fig.~\ref{fig:sketch-chips}): the distance~$d$ is measured from the beam position to the centre of the neighbouring wire.
Grating (Sec.\,\ref{s:mag-grating}, Fig.~\ref{fig:wire-grating}): the distance~$y$ is measured from the top wire surface; the wire cross-sections 
are~$\rm40\times2\,\mu m^2$ and their centre-to-centre separation is~$\rm50\,\mu m$.
The splittings are estimated for a~$\rm0.1\,eV$ ($\rm700\,m/s$) beam with~$\rm0.2\,mrad$ divergence and the emittance of Table~\ref{t:source} (left column).
}
\label{t:chip-values}
\item
\begin{tabular}{lll}
\br
                                       & wire pair 	                           & grating                               \\
\mr
current		                             & $I=\rm100\,mA$ 	                     & $\rm1\, A$ 	                         \\
cross section                          & $\rm2\times0.5\,\mu m^2$  	           & $\rm40\times2\,\mu m^2$               \\
current density                        & $\rm10^7\,A/cm^2$   	                 & $\rm1.25\cdot10^6\,A/cm^2$            \\
distance	                             & $d=\rm1\,\mu m$                       & $y\ge\rm20\,\mu m$                    \\
length		                             & $L=\rm100\,\mu m$ 	                   & $\rm20\,mm$                           \\
mag. gradient		                       & $B'=\rm4.0\times10^4\,T/m$            & $\rm640\,T/m$                         \\
SG splitting		                       & $\Delta v^{\rm SG}_\perp/v=\rm2.1\,mrad$   	   & $\rm11\,mrad$               \\
broadening		                         & $\delta v_\perp/v=\rm0.32\,mrad^{(a)}$          & $\rm5\ldots18\,mrad^{(b)}$  \\
\br
\end{tabular}%
\item[$\rm^{(a)}$ ] Due to the inhomogeneous Lorentz force.
\item[$\rm^{(b)}$ ] Due to a range of penetration depths into the inhomogeneous field, depending on the angular and velocity spreads (data from Table~\ref{t:source}). Smaller value: ion beam focused on the reflection point [Fig.~\ref{fig:two-blobs}(\emph{c})]. 
\end{indented}
\end{table}

An ion beam that travels with velocity~$v$ parallel to the wires will be split transversally, if the spin is polarised perpendicular to the beam axis. The inhomogeneous field generates a force that acts differently on the two spin states during the ion flight time through the wire gap, leading to an angular splitting~$\Delta\theta\simeq\Delta v_\perp/v^{\rm SG}$ between the two spin states, where 
\begin{equation}
\Delta v_\perp^{\rm SG} \approx 
	\frac{ L }{ m v }\ (\bfmu \cdot \nabla) {\bf B}
	\sim
    \frac{ g_e \mu_B \Delta S L }{ m v }
		\
    \frac{ \mu_0 I }{ \pi d^2 }
\label{eq:estimate-split-0}
\end{equation}
Over an interaction length~$L=\rm100\,\mu m$, values of~$\Delta\theta\gtrsim\rm2\,mrad$ can be achieved, provided the incident beam is sufficiently slow, 
$v<\rm700\,m/s$ (i.e., beam energy~$<\rm0.1\,eV$ for the~$\rm^{40}Ca^+$ isotope), with~$m=\rm39.96\,amu$ and~$\Delta S=1$. This is a significantly larger separation than our estimate for the magnetic wedges, see Eq.(\ref{eq:mag-pole-splitting}).

We note that Eqs.(\ref{eq:estimate-gradB}--\ref{eq:estimate-split-0}) are based on the approximation of infinitely thin wires. Numerical calculations accounting for the finite width and thickness of the wires yield somewhat higher gradients for gaps that are narrower than the wires. These are the data used for Fig.~\ref{fig:sketch-chips}(\emph{d}).

An experimental realisation using this configuration should separate spin components of the beam in the~$\pm y$-direction (perpendicular to the chip surface) in order to avoid having the beam ``crash'' into the ``side walls'' of the gap that is only~$\rm0.1\,\mu m$ wide. This corresponds to initial spin states along the~$\pm x$-axis [see Eq.(\ref{eq:SG-force-quadrupole})]. By adding a homogeneous field along the~$x$-axis, the adiabatic behaviour of the spin as it enters the region between the two wires can be maintained.

\paragraph{Differential Lorentz forces.}

To estimate broadening due to the Lorentz force, we start from the ion beam data (see Table~\ref{t:source} and Ref.~\cite{Jacob2016}), in particular the source emittance of~$\eta=\rm1.6\,nm\,mrad\sqrt{eV}$ (in~1D; this value is the square root of the~2D emittance). For the required beam energy of~$E=\rm0.1\,eV$ and a beam divergence of~$\Delta\theta_y\simeq\rm0.2\,mrad$, this results in a relatively narrow beam waist of size~$\Delta y=\eta/(\sqrt{E}\,\Delta\theta_y)\simeq\rm25\,nm$. The magnetic field gradient translates this into a transverse velocity spread due to the variation of the Lorentz force, in the same direction as the~SG splitting,
\begin{equation}
\delta v^{\rm L}_\perp \sim
\frac{ L }{ m }\ e B' \, \Delta y 
\,.
\label{eq:v_perp-Lorentz}
\end{equation}
From the magnetic gradient introduced in Eq.(\ref{eq:estimate-gradB}) above, we find an angular broadening of~$\sim\rm0.32\,mrad$, much less than the angular splitting from the~SG force (Table~\ref{t:chip-values}). We therefore conclude that the splitting should be observable, although the small dimensions of the gap between the wires appear challenging for the ion beam optics and similarly challenging for the nano-fabrication of the long and narrow, high aspect ratio channel.

Note that in these estimates, the change in transverse velocity is assumed to be proportional to the flight time through the inhomogeneous field, ignoring the possibility of an oscillating force. This may occur, however, if the beam path is bent or when precession changes the spin direction relative to the magnetic gradient. This issue will be considered in the following section.


\subsection{Wire grating}
\label{s:mag-grating}

\begin{figure}[t]
\vspace*{2ex}
\centerline{%
\includegraphics*[width = 0.6\textwidth]{\figpath 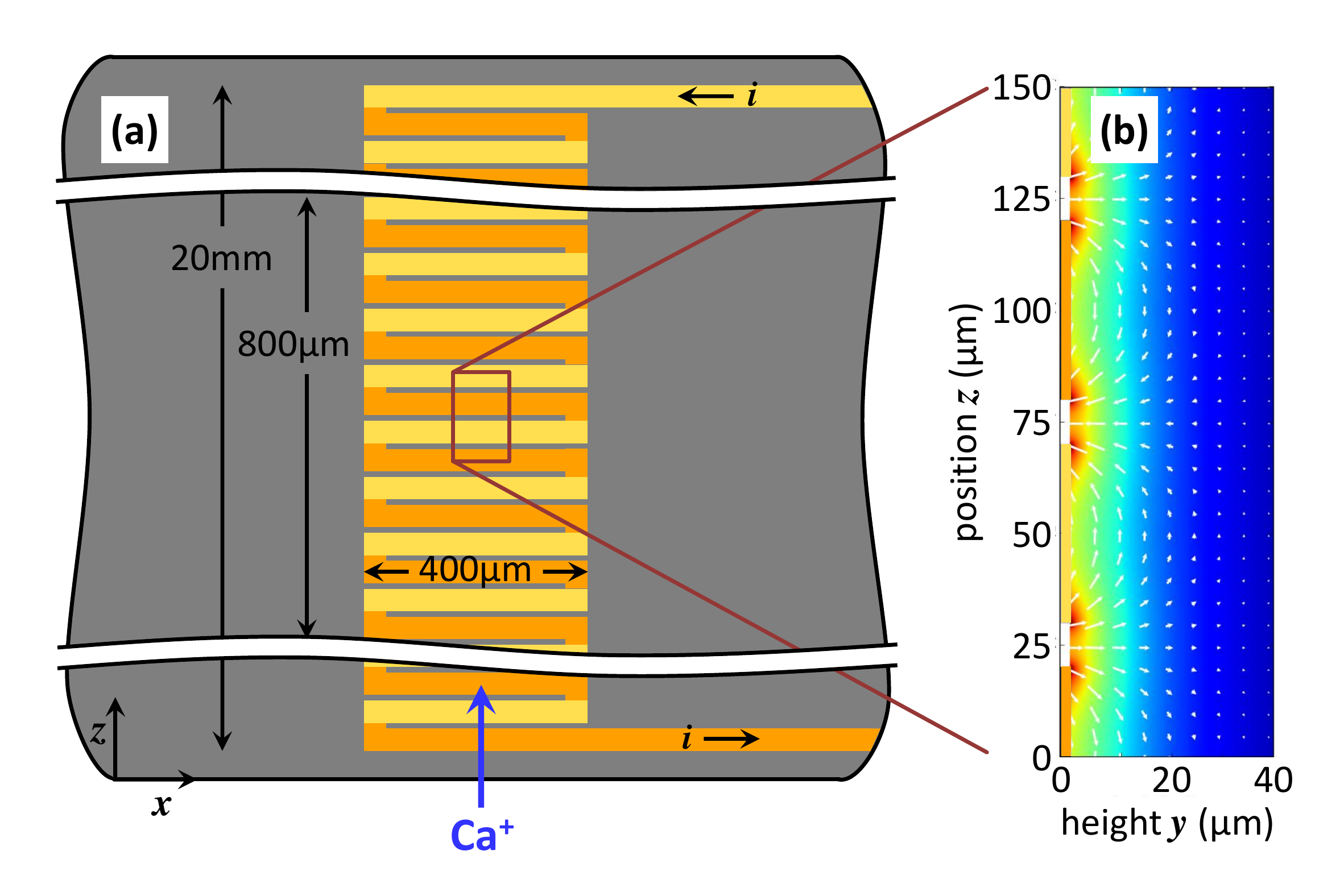}%
}
\vspace*{-3ex}
\caption[]{Magnetic field created by an array of wires. 
(\emph{a})~Implementation on an atom chip with~DC currents in alternating directions (dark/light gold-coloured wires). The~$\rm2\,\mu m$-thick wires 
are~$\rm40\,\mu m$ wide and~$\rm400\,\mu m$ long, while the gaps are~$\rm10\,\mu m$ wide. The eight wire pairs shown in the central section of the chip have a 
combined length of~$\rm800\,\mu m$. The entire chip would have~200 wire pairs, for a combined length of~$\rm20\,mm$, as shown. 
(\emph{b})~Cut through the wires shown in~(\emph{a}), displaying the direction (arrows) and magnitude (colours, arbitrary units) of the magnetic field above the top surface of the wires.
}
\label{fig:wire-grating}
\end{figure}

Here we take advantage of state-of-the-art micro-chip fabrication techniques~\cite{Keil2016}. We consider using a set of parallel wires, as sketched in Fig.~\ref{fig:wire-grating}, to create a magnetic field that is localised near the chip surface. This can be built with a very large number of wires~\cite{Cognet1999}, each with a cross-section of order~$\rm100\,\mu m^2$. A similar field pattern may also be generated by magnetised microstructures rather than electric currents~\cite{Roach1995,Hinds1999b}, thereby avoiding technical problems related to Joule heating. For simplicity however, we focus on the periodic wire array in the following discussion.

\begin{figure}[t]
\centerline{%
\includegraphics*[trim=0 15mm 0 15mm,clip,width = 1.1\textwidth]{\figpath 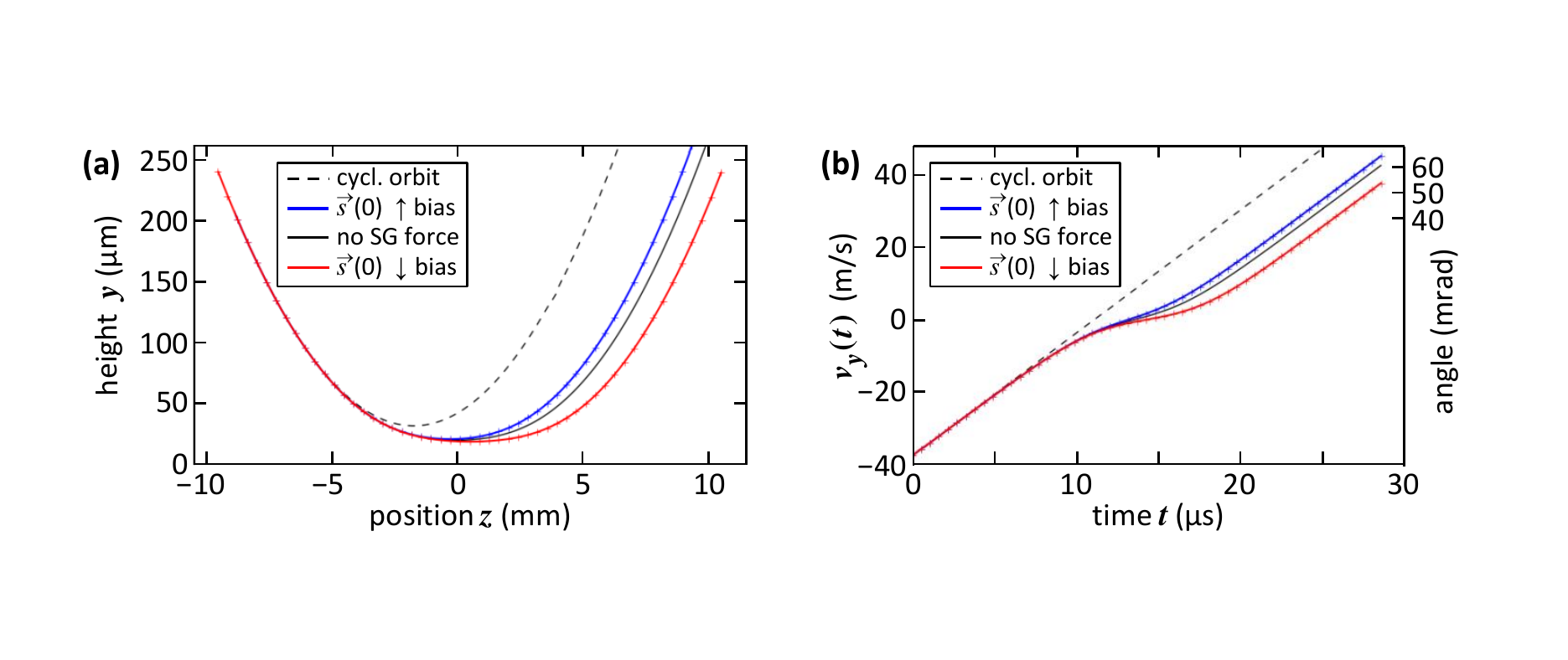}
}
\caption[]{
Trajectories flying above the wire grating sketched in Fig.~\ref{fig:wire-grating}.
(\emph{a})~Projection onto the plane of incidence, perpendicular to the wires. 
(\emph{b})~Vertical velocity vs.\ flight time. Solid lines: semiclassical calculation including the spin precession in the inhomogeneous field. Symbols:  adiabatic approximation based on Eq.\,(\ref{eq:slow-vertical-force}). The dashed lines in~(\emph{a}) and~(\emph{b}) illustrate the cyclotron orbit. The beam deviates from it because of the attractive image force. The  splitting from the spin-dependent force occurs around the same time. Spins are launched parallel or antiparallel to the bias field~[$\vec s(0)\!\!\!\uparrow$ and~$\vec s(0)\!\!\!\downarrow$\,, respectively]. In addition to the parameters given in
Table~\ref{t:chip-values}, we consider field strengths~$B_0=\rm20\,G$ [the `transverse bias' of Eq.\,(\ref{eq:avoid-the-crash}), perpendicular to the figure plane] and~$B_1\approx\rm360\,G$ [corresponding to the coefficient~$A_1$ in Eq.\,(\ref{eq:expansion-Phi}) and measured at the top wire surface], 
i.e.~$\approx\rm102\,G$ at~$y=\rm20\,\mu m$ distance. The ions impinge on the grating at~$\rm700\,m/s$ and grazing incidence~($\approx\rm54\,mrad$). An interaction distance of~$\rm20\,mm$ corresponds to~200 wire pairs and a flight time of~$\rm29\,\mu s$.
}
\label{fig:bent-paths}
\end{figure}

The ion beam would be incident in the~$yz$-plane at a grazing angle, nearly parallel to the~$z$-axis [blue arrow in Fig.~\ref{fig:wire-grating}(\emph{a})]. As we move with an ion along a path at constant height~$y=y_0$, the magnetic field rotates in the~$yz$-plane, see Fig.~\ref{fig:wire-grating}(\emph{b}). The field gradient in the~$y$-direction leads to a spin-dependent force in this direction. By solving the equations of motion, we find a non-zero~SG force despite the oscillating field (see Fig.~\ref{fig:bent-paths}). A similar concept has been demonstrated by Bloom and co-workers and named the ``transverse~SG force''~\cite{Bloom1962,Bloom1967}.

The magnetic potential generated by the set of wires in the upper half-plane can be written as a Fourier series 
\begin{equation}
\Phi( y, z ) = - \sum_{n \ge 1,\,{\rm odd}}^{\infty} 
A_n\, {\rm e}^{ - n \kappa y }\sin( n \kappa z )
\,,\qquad
y \ge 0
\label{eq:expansion-Phi}
\end{equation}
where~$\pi/\kappa$ is the distance between neighbouring wire centres (with opposite currents). The coefficients of the exponentially decaying terms are
\begin{equation}
A_n = \frac{ 2 \mu_0 I }{ \pi n }\
\frac{ \sin( n \kappa w / 2 ) }{ n \kappa w }\
\frac{ 1 - {\rm e}^{ - n \kappa t } }{ n \kappa t }
\,,\qquad
n \ne 0 \,.
\label{eq:Bn-coefficients}
\end{equation}
Here, the wire cross sections are taken as rectangular with width~$w$ and thickness~$t$; $I$ is the current per wire, with a homogeneous distribution. 
One gets this result by expanding the current density and the potential in a Fourier series and solving for its Fourier coefficients inside and outside 
the layer~$-t<y<0$.


\paragraph{Image potential and transverse bias.}

Electric fields have to be avoided along the ion path as they generate unwanted forces. It has already been mentioned by Enga and Bloom~\cite{Enga1970} that  voltage drops along the current-carrying wires generate sizeable electric forces. They can be shielded by covering the wire array with a grounded conducting layer, using indium tin oxide for example. Still, the nearby ion will induce surface charges in this conducting layer. The corresponding force can be computed from the image potential, which is half the Coulomb potential of a symmetrically placed charge below the layer:
\begin{equation}
V_{\rm im}( {\bf r} ) =
- \frac{ e^2 }{ 16\pi\varepsilon_0 y }
\,,
\label{eq:image-potential}
\end{equation}
where~$y=0$ is the position of the surface. (We neglect the retarded response of the image charge due to the motion of the ion, the light roundtrip time, and the delayed response of the surface. All these effects lead to a friction force~\cite{VolokitinPerssonBook}.) At a typical distance of~$y=\rm10\,\mu m$, the image potential corresponds to~$\sim\rm36\,\mu eV$ which can be comparable to the transverse kinetic energy of a low-energy beam. One cannot avoid getting relatively close to the surface (of the order of~$1/\kappa\approx\rm15\,\mu m$) because of the rapid decay of the magnetic field. 

The image force attracts the ion to the surface. One way to compensate for this force and avoid having the beam ``crash'' into the surface would be to generate a repulsive electric field by applying a voltage to a finite portion of the covering layer. An alternative concept (that we elaborate on here) is to use a transverse magnetic bias field, oriented along the~$x$-axis parallel to the wires. The sign of this bias is chosen such that the cyclotron orbits are 
`bending upwards', as illustrated in Fig.~\ref{fig:bent-paths}. The required bias field~$B_0$ is determined by the inequality
\begin{equation}
e v_z B_0 > \frac{ e^2 }{ 16\pi\varepsilon_0 \, y^2 }
\label{eq:avoid-the-crash}
\end{equation}
for all values of~$y$ along the trajectory, and falls in the~$\rm10\ldots30\,G$ range for the typical values of~$v_{z}$ (ion velocity) and~$y$ (distance)
adopted here. The bias also serves as a quantisation axis for the spin when the incoming ions are far from the magnetic grating.

\paragraph{Ion trajectories.}
We have solved numerically the coupled equations of motion for the spin and the centre of mass of the ion [\ref{a:averaged-SG}, 
Eqs.(\ref{eq:spin-precession}--\ref{eq:eqn-motion})]. The beam does not deviate more than about~$\rm10\,nm$ from the plane of incidence~($yz$-plane). The~SG force for this configuration is a `transverse' one~\cite{Bloom1962,Bloom1967}, since the spin starts aligned with the bias field~($x$-direction), perpendicular to the magnetic gradient~($y$-direction). We discuss in~\ref{a:averaged-SG} how the spin acquires oscillating components in the~$yz$-plane that generate a spin-dependent force with a nonzero average.

If one considers, as in the standard~SG setup, that all fields are confined to some region, say the one shown in Fig.~\ref{fig:bent-paths}(\emph{a}), then we have the following scenario. The ion beam approaches the chip surface~($y=0$ in the figure) at a glancing angle. It performs a semi-circular cyclotron orbit 
as it enters the bias field (dashed curves). It drifts off this orbit upon approaching the chip because of the attractive image force. If the 
condition~(\ref{eq:avoid-the-crash}) is met, the beam will eventually bend away from the chip and leave the region of the magnetic field. During the phase of `closest approach' (about~$\rm20\,\mu m$ from the chip surface), the spin-dependent force splits the two spin states (blue and red trajectories with symbols). When the ion leaves the grating, the splitting amounts to an angular separation close to~$\rm10\,mrad$ and a spatial separation of more than~$\rm50\,\mu m$ at the exit of the interaction region considered here, see Fig.~\ref{fig:bent-paths}(\emph{b}). The spin-dependent splitting can be simulated with semi-classical trajectories by launching the spin in a suitable state. Our semiclassical calculation finds that the largest splitting occurs when, far from the grating, the spin is aligned parallel or antiparallel to the bias field, as expected for adiabatic behaviour. Our results are consistent with quantum-mechanical simulations that evolve a two-component wave function. A more thorough analysis and comparison will be provided in a separate paper~\cite{next_theory_paper}.

Let us note that there are three characteristic frequencies in the problem: the largest one is the Larmor frequency of spin 
precession,~$\Omega\approx g\mu_B|{\bf B}|/\hbar\approx\rm2\pi\times2.8\,MHz\times(|{\bf B}|/G)$. For typical fields of~$\rm20-100\,G$ considered here, this is in the range of~$\rm56-280\,MHz$. Next is the rotation frequency~$\kappa v_z/2\pi\approx\rm7\,MHz$ of the magnetic field in the frame co-moving with the beam. We show in~\ref{a:averaged-SG} that the rotating field can be removed with a suitable coordinate transformation; then adiabaticity holds if the direction of an effective magnetic field defined in Eq.\,(\ref{eq:Beff-rotating-frame}) changes slowly enough. We find that this condition is reasonably well satisfied for the parameters chosen here. The slowest frequency scale is~$\omega=e|{\bf B}|/m\approx\rm2\pi\times0.8\ldots3.8\,kHz$ that generates the cyclotron orbits. 

It is also demonstrated in \ref{a:averaged-SG} that a net~SG force arises for a `spinning charge' although the spin precession is fast. The upshot is that in the vertical direction, one gets an averaged (or adiabatic)~SG force whose sign can be determined by the initial spin polarisation along the bias field. We find the approximate acceleration
\begin{equation}
\frac{ F_y }{ m } \approx \omega_0 v_{z}
- \frac{ e^2 }{ 16\pi\varepsilon_0 m y^2 }
+ \frac{ \omega_1^2\,{\rm e}^{-2\kappa y} }{ 2 \kappa }
+
	u \omega_1\  
	\frac{ \Omega_1 \,{\rm e}^{-2\kappa y} }{
    \tilde \Omega(y)}\ 
    S_{x0}
\,.
\label{eq:slow-vertical-force}
\end{equation}
Here~$\omega_{0,1}$ and~$\Omega_{0,1}$ are the cyclotron and Larmor frequencies corresponding to~$B_0$ and~$B_1$ [the latter is the field corresponding to the coefficient~$A_1$ in Eq.\,(\ref{eq:expansion-Phi})], while $u\approx\hbar\kappa/m_e$ is a characteristic velocity proportional to the grating vector~$\kappa$ [see Eq.\,(\ref{eq:force-z})], and~$\tilde\Omega$ is given by Eq.\,(\ref{eq:Beff-rotating-frame}):
\begin{equation} 
\tilde \Omega( y ) 
= [(\Omega_0 - \kappa v_{z0})^2 + \Omega_1^2 \,{\rm e}^{-2\kappa y}]^{1/2}
\,.
\label{eq:def-Omega-of-y}
\end{equation}
Finally,~$S_{x0}$ is the spin projection onto the magnetic field far from the surface; for the trajectories shown here,~$S_{x0}=\pm\frac12$. The first two terms in Eq.\,(\ref{eq:slow-vertical-force}) correspond to the Lorentz force and the image force respectively. The third term may be called a ponderomotive force, and arises from the `wiggling' of the ion in the oscillating magnetic field (as in an undulator). The last term is the~SG force averaged over the spin precession. The trajectories resulting from this approximation are shown by the symbols and agree very well with the full numerical solution shown as solid lines in Fig.~\ref{fig:bent-paths}. 

We find splittings between the spin states that can be even larger than the~$\rm10\,mrad$ shown here when the ion penetrates down to a~$\rm15\,\mu m$ distance from the top wire surface (corresponding to a larger angle between the surface and the incident beam). The transverse bias field would then have to be increased to~$>\rm25\,G$ to compensate for the image force [see~Eq.(\ref{eq:avoid-the-crash})]. Large splittings of such a size can be understood from the interplay between the~SG force on the one hand, and the Lorentz and image forces on the other. From Eq.(\ref{eq:slow-vertical-force}), one may construct two approximate potentials that govern the vertical motion for the spin states which are initially eigenstates of the spin operator along the transverse bias field with eigenvalues~$S_{x0}=\pm\frac12$. Neglecting the small changes in the axial velocity~$v_z$, these two potentials lead to trajectories with different turning points due to contributions of the attractive or repulsive~SG forces for the corresponding spin states: the state subject to an attractive~SG force penetrates closer to the surface than the other spin state. This creates a delay between the two trajectories when they leave the surface while being accelerated away by the Lorentz force, so that at a given time and position along~$z$, they have different vertical velocities and hence they are angularly split. More details will be discussed in Ref.~\cite{next_theory_paper}. 

We note that the spatial separation of the spin states [some tens of microns, see Fig.~\ref{fig:bent-paths}(\emph{a})] is large enough that one may build a spin filter for a particular spin polarisation by placing slits at a suitable position downstream from the magnetic grating. One could also restrict the magnetic bias field to a finite region along the beam in order to have straight trajectories rather than cyclotron orbits once the spin states are split.

\begin{figure}[t]
\centerline{%
\includegraphics*[width = 1.0\textwidth]{\figpath 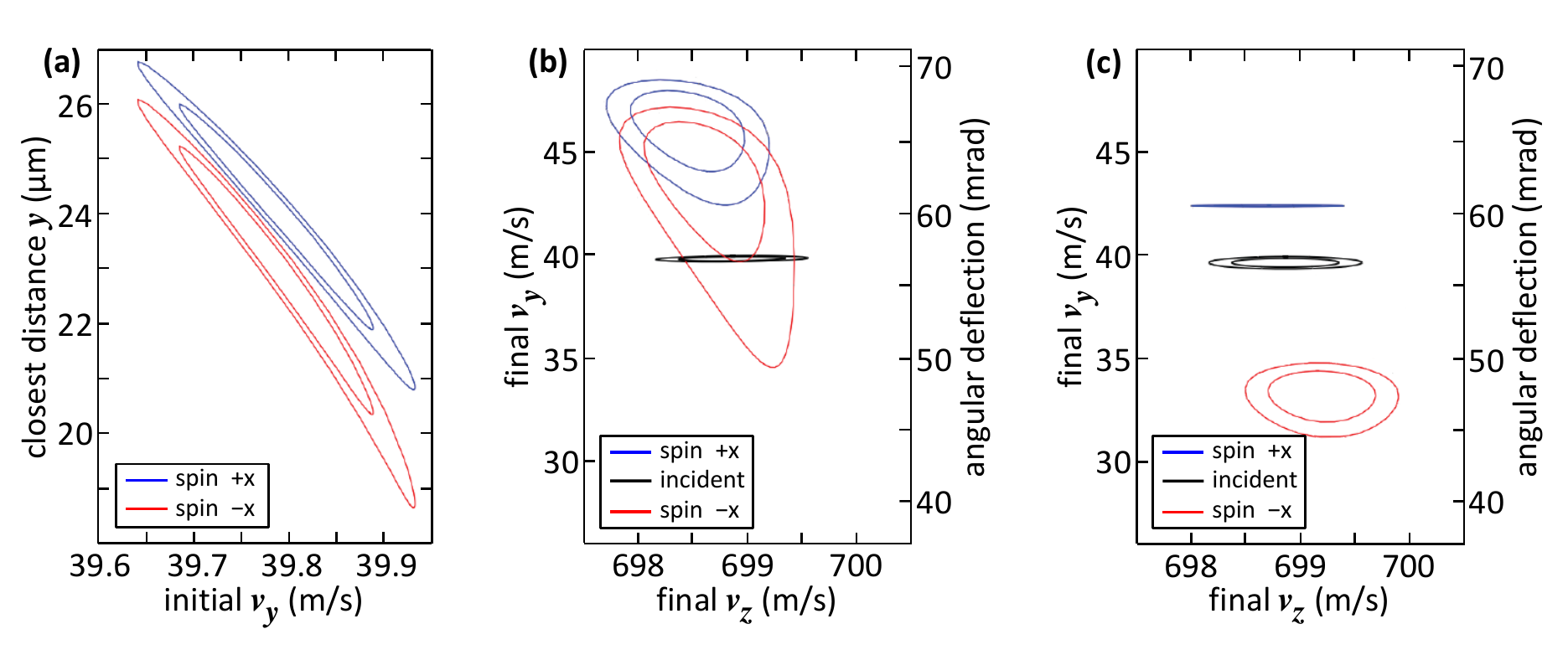}
}
\caption[]{%
Broadening of the velocity distribution upon reflection and splitting from the magnetic grating. We show contours of ion distributions
computed by allowing for spreads in the beam velocity and angle. The outer contours use the velocity and angular spreads in Table~\ref{t:source}, 
second column (the product~$\delta v\,\delta\theta$ is very similar for both columns), while the inner contours use~30\% narrower spreads for both variables.
(\emph{a}) The physical origin of the broadening is that the distance of closest approach depends on the initial vertical velocity. Both the~SG and the image force reach their highest values at close distances. The spread in this distance increases the angular spread of the reflected beam. As noted in the text, these forces are much larger than the Lorentz force in this configuration. The upper and lower contours correspond to spin components: they are already split when reaching the closest distance because of the opposite sign of the~SG force.
(\emph{b}) Distribution of final velocities after a flight path of~$\rm20\,mm$ over the magnetic grating. $v_z$:~parallel to the chip surface, 
$v_y$:~perpendicular. The right-hand axis gives the deflection angle in~mrad. The narrow black contour illustrates the incident beam.
(\emph{c}) Same as (\emph{b}) but the initial conditions are chosen such that the beam reaches a distance of closest approach of~$\rm20\pm0.25\,\mu m$. The beam energy spread is the same.
}
\label{fig:two-blobs}
\end{figure}

\paragraph{Broadenings.}
The angular distribution of the ion beam is broadened due to the range of penetration depths into the field of the magnetic grating, and this effect overwhelms the broadening due to the spatially inhomogeneous Lorentz force that we considered in the two-wire case. The distance of closest approach is determined by the incident velocity component~$-v_{y0}$ (normal to the surface) and the spin state~$S_{x0}$. The ions that approach closest are subject to the largest forces. The contours in Fig.~\ref{fig:two-blobs} illustrate distributions for different pairs of parameters. In panel~(\emph{a}), we plot the distance of closest approach vs.\ the normal velocity~$-v_{y0}$ and note a relatively wide range of distances. The corresponding angular broadening is shown in panel~(\emph{b}): it is larger than the splitting between the spin states and would therefore prevent their resolution. However, due to the very high precision experimentally realised by the ion optics~\cite{Jacob2016}, we may focus the beam onto the point of closest approach. The result is illustrated in panel~(\emph{c}), where a clean~SG splitting is visible. 

Here, the initial distribution of positions and angles is adjusted so that the closest distance falls into a~$\rm\pm250\,nm$ wide range, indicated by the outer black ellipses in~(\emph{b, c}). This range was taken rather arbitrarily; the ion source is actually capable of producing a narrower focus, as demonstrated experimentally~\cite{schnitzler2009deterministic,Jacob2016} at a higher beam energy. We therefore consider the wire grating to be the practical configuration that would be most likely to enable charged-particle spin separation and related experiments.

\section*{Conclusion}

We have presented three proposals for experimental setups enabling the Stern-Gerlach~(SG) effect with an ion beam to be observed. Based on state-of-the-art magnet~\cite{Hsu2016} and fabrication technology~\cite{Keil2016}, and parameters from experiment~\cite{Jacob2016}, we conclude that with a slow beam having a kinetic energy of~$\approx\rm0.1\,eV$, all three configurations show observable spin-dependent splittings. The main difference from electrons, where~SG splitting is only marginally possible~\cite{Batelaan2002,Garraway2002}, is that ions are much more massive and therefore the spread due to spatially inhomogeneous Lorentz forces does not pose a fundamental problem. 

The configuration of Sec.\,\ref{s:mag-poles} with two magnetic poles is closest to the original~SG setup and has the advantage of a relatively simple design. We showed [Fig.\,\ref{fig:mag-wedge}(\emph{d})] that properly focused trajectories and magnetic pole design can lead to spin-dependent separations even with strong bending due to the Lorentz force. The other two configurations rely on atom chip fabrication. A simple two-wire configuration (Section\,\ref{s:wire-pair}) shows sufficient~SG splitting to be resolvable even in the presence of Lorentz bending and broadening (Table~\ref{t:chip-values}).  It is however, a difficult structure to fabricate because of the high aspect ratio required, and it would also pose a real ion optics challenge, as the trajectories have to be `threaded' through very narrow gaps (in the~$\rm0.1\,\mu m$ range). Finally, the periodic magnetic field of Sec.\,\ref{s:mag-grating} has the advantage of an open, planar design and exhibits the largest angular spin splitting~($\sim\rm10\,mrad$). This is a factor of~$\sim50$ larger than the design value of the ion beam divergence. The strong ion-surface forces that appear at~$\sim\rm20\,\mu m$ distance (attractive image charge) can be mitigated by adding a homogeneous magnetic field. This field has the additional advantage of defining a quantisation axis for the ion spin far from the grating. Although the particles cross a large number of grating periods, the~SG force accumulates over time, similar to the `transverse~SG effect' \cite{Bloom1962,Bloom1967}. The main challenge in implementing this configuration is to control the spreading of the distance of closest approach; we show that this can be done by carefully focusing the ion beam at the turning point above the magnetic grating. 

According to our analysis, inhomogeneous magnetic fields can be used to separate spin states of a charged particle, thereby also enabling a spin filter. We did not extend this analysis to include the coherence of the spin-dependent splitting, which would require  careful control of the magnetic fields that determine the relative phase between the spin states. This has been demonstrated in a different setup with neutral atoms~\cite{Machluf2013,Margalit2018}, despite earlier claims in the literature that this would not be possible for the~SG effect~\cite{Englert1997}. The proposal presented here extends these exciting possibilities to ion beams that have a strong potential for sensing and can be manipulated with excellent experimental control.

Given the $\rm100^{th}$ anniversary of work started by Stern and Gerlach in their Frankfurt laboratory, we present this paper in honour of those first heroic efforts, and hope that it may open a road for new steps in this fundamental story.


\paragraph{Acknowledgments.}
We thank Ulrich Poschinger for helpful discussions. This work has been supported by the \emph{Deutsche Forschungsgemeinschaft} through the~DIP program (grant nos.~Schm-1049/7-1 and Fo 703/2-1). It is funded in part by the Israel Science Foundation (grant no. 856/18). CH~is grateful to~RF, YJ, and~MK for hospitality at Ben Gurion University where this work was completed. We also acknowledge support from the Israeli Ministry of Immigrant Absorption (to~MK).


\appendix

\section{Period-averaged Stern-Gerlach force}
\label{a:averaged-SG}

The trajectories shown in Sec.\,\ref{s:mag-grating} are based on solving semiclassical equations of motion including spin precession, as discussed in 
Refs.~\cite{JacksonBook,McGregor2011}. One uses Eq.\,(\ref{eq:g-factor}) for the magnetic moment and works with the expectation value~${\bf S}$ of the spin operator:
\begin{eqnarray}
\frac{ {\rm d} {\bf S} }{ {\rm d}t } &=&
- \frac{ g_e \mu_B }{ \hbar }\ {\bf S} \times {\bf B}( {\bf r} )
\label{eq:spin-precession}
\\
m \frac{ {\rm d} {\bf v} }{ {\rm d}t } &=&
- g_e \mu_B ({\bf S} \cdot \nabla) {\bf B}( {\bf r} )
+
e {\bf v} \times {\bf B}( {\bf r} )
-
\nabla V_{\rm im}( {\bf r} )
\,.
\label{eq:eqn-motion}
\end{eqnarray}
Here~$e$ is the ion charge, $V_{\rm im}({\bf r})$ is the image potential~(\ref{eq:image-potential}), and we have again used~$\nabla\times{\bf B}={\bf0}$. 
The approximation behind these equations of motion is that products of spin and centre-of-mass observables (e.g., velocities~{\bf v}) factorise; it can be improved by tracking the corresponding correlation functions, e.g. $C_{Sv}=\langle\hat{\bf S}\hat{\bf v}\rangle-{\bf S}{\bf v}$. It is obvious, however, that one can describe a~SG splitting at this semiclassical level, since the force~(\ref{eq:eqn-motion}) depends on the spin orientation. This is illustrated by the upper and lower curves in Fig.~\ref{fig:bent-paths} which originate from two opposite initial spin directions.

In the following, we specialise to a magnetic field of the form
\begin{equation}
{\bf B} = B_0 \hat{\bf x} +
B_1( y ) 
\left(
\hat{\bf z} \cos \kappa z - 
\hat{\bf y} \sin \kappa z
\right)
\,,\qquad
B_1( y ) =
B_1 \,{\rm e}^{ - \kappa y } 
\label{eq:spin-lateral-bias}
\end{equation}
where we keep for simplicity only the first term of the Fourier expansion~(\ref{eq:expansion-Phi}). At distances comparable to the grating 
pitch~$2\pi/\kappa$, the other terms will be much smaller. The equations of motion for the spin~(\ref{eq:spin-precession}) are written in components
\begin{eqnarray}
\frac{ {\rm d} S_x }{ {\rm d}t } &=& 
- \Omega_1( y ) S_y \cos \kappa z
- \Omega_1( y ) S_z \sin \kappa z
\label{eq:spin-x}
\\
\frac{ {\rm d} S_y }{ {\rm d}t } &=& 
- \Omega_0 S_z
+ 
\Omega_1( y ) S_x \cos \kappa z
\label{eq:spin-y}
\\
\frac{ {\rm d} S_z }{ {\rm d}t } &=& 
\Omega_0 S_y 
+ 
\Omega_1( y ) S_x \sin \kappa z
\label{eq:spin-z}
\end{eqnarray}
where~$\Omega_n=g_e\mu_B B_n/\hbar$ ($n=0,1$) are the spin Larmor frequencies. The centre-of-mass motion follows from the forces in the magnetic 
field~(\ref{eq:spin-lateral-bias}). The interesting part is the modulated field because it determines the~SG force 
\begin{equation}
{\bf F}_{\rm SG} = g_e \mu_B \kappa B_1( y )
\left[
\hat{\bf y} (S_z \cos \kappa z
-
S_y \sin \kappa z)
+
\hat{\bf z} (S_y \cos \kappa z \,.
+
S_z \sin \kappa z)
\right]
\label{eq:xSG-force}
\end{equation}
The Lorentz force~$e{\bf v}\times{\bf B}$ defines the cyclotron frequencies~$\omega_0=eB_0/m$ and~$\omega_1(y)=eB_1(y)/m$ where~$m$ is the ion mass. Putting this together, the equations of motion are
\begin{eqnarray}
\frac{ {\rm d}v_x }{ {\rm d}t } &=&
\omega_1( y ) v_y \cos \kappa z
	+ \omega_1( y ) 
	   v_z \sin \kappa z
\label{eq:force-x}
\\
\frac{ {\rm d}v_y }{ {\rm d}t } &=&
\omega_0 v_z
+ a_{\rm im}(y)
- \omega_1( y ) v_x \cos \kappa z
+ \omega_1( y ) 
u (S_z \cos \kappa z
	- S_y \sin \kappa z )
\label{eq:force-y}
\\
\frac{ {\rm d}v_z }{ {\rm d}t } &=&
- \omega_0 v_y
- \omega_1( y ) v_x \sin \kappa z 
+ \omega_1( y ) u
  (S_y \cos \kappa z
  +
  S_z \sin \kappa z)
\label{eq:force-z}
\end{eqnarray}
where~$a_{\rm im}$ is the acceleration due to the image potential and~$u=g_e\mu_B\kappa/e$ is the characteristic velocity for the~SG force, of the order of the `recoil velocity' of an electron~$\hbar\kappa/m_e\approx\rm7.3\,m/s$ for the wire array considered in Table~\ref{t:chip-values}. The numerical solution of these equations gives the trajectories shown in Fig.\,\ref{fig:bent-paths} (solid lines). In the following, we focus on short time scales where the distance~$y$ is varying slowly, where we aim to simplify the magnetic forces by taking averages over the grating period.


\paragraph{Analysis: oscillating vs period-averaged.}

Let us assume that the beam is mainly moving along the~$z$-direction with a fixed velocity~$v_{z0}$. We assume that the transverse components are small initially and remain small (grazing incidence). As noted in the main text, there are three frequency scales in the problem: the spin precession (Larmor) frequencies~$\Omega_{0,1}$ involve the gyromagnetic ratio $g_e\mu_B/\hbar\approx2\pi\times2.8\,{\rm MHz/G}$. They are typically above~$\rm10\,MHz$, comparable or larger than the `Doppler' frequency~$\kappa v_{z0}$. The Lorentz force translates into much lower frequencies because 
$\omega_{0,1}\approx2\pi\times 38\,{\rm Hz}\,(B_{0,1}/{\rm G})$. The corresponding cyclotron radius~($\sim 0.1\,{\rm m}$) is much larger than the other characteristic
distances.

For the spin, the dynamics is very similar to the transverse spin resonance problem; it is best analysed in a frame co-moving along the beam and rotating at the frequency~$\kappa v_{z0}$ around the~$x$-axis (parallel to the bias field). In this frame (marked by the primes), the magnetic field becomes static, 
\begin{equation}
{\bf B}(y, v_{z0}t) 
=
\mbox{\boldmath${\sf R}$}_x(\varphi_t) 
	{\bf B}'(y)
=
\mbox{\boldmath${\sf R}$}_x(\varphi_t) 
	[ B_0 \hat{\bf x} + B_1(y) \hat{\bf z} ]
\label{eq:rotate-B}
\end{equation}
where $\mbox{\boldmath${\sf R}$}_x(\varphi_t)$ is the rotation matrix around the axis~$x$ with angle~$\varphi_t=\kappa v_{z0}t$. The equation of motion for the spin~${\bf S}'(t)$ describes a precession in this frame with frequency~$\tilde\Omega(y)$ around a fixed axis given by the unit vector~$\hat{\bf n}'$
\begin{equation}
\hat{\bf n}' = 
[ \Omega_0' \hat{\bf x} + \Omega_1(y) \hat{\bf z} ]
/ \tilde \Omega( y ) 
\,,\qquad
\tilde \Omega( y ) 
= [\Omega_0^{\prime 2} + \Omega_1(y)^2]^{1/2}
\label{eq:Beff-rotating-frame}
\end{equation}
Note the `Doppler shift'~$\Omega_0'=\Omega_0-\kappa v_{z0}$ that makes this axis deviate from the naive alignment parallel to~${\bf B}'$ [Eq.\,(\ref{eq:rotate-B})]. The composition of the two rotations, 
$\mbox{\boldmath${\sf R}$}_x(\varphi_t) 
 \mbox{\boldmath${\sf R}$}_{{\bf n}'}(\tilde\varphi_t)$
with~$\tilde\varphi_t=\tilde\Omega t$, generates sum frequencies that render the spin dynamics relatively complex. We now apply the adiabatic approximation: 
$\tilde\Omega$ is assumed to be the largest frequency so that the spin remains aligned (anti)parallel to the precession axis~(\ref{eq:Beff-rotating-frame}):
\begin{equation}
{\bf S}'(t) = \hat{\bf n}'(y(t)) 
\big[ \hat{\bf n}' \cdot {\bf S}' \big]
\,.
\label{eq:adiabatic-spin}
\end{equation}
We have checked that the scalar product appearing here is indeed constant in time to a good approximation. It can thus be evaluated at the initial stage of the trajectory where the direction~$\hat{\bf n}'$ is parallel to the bias field: we then get~$\hat{\bf n}'\cdot{\bf S}'=S_{x0}=\pm\frac12$, the initial spin projection.

In the adiabatic approximation, the centre-of-mass motion evolves more slowly than the spin. For the~SG force~[last term in Eq.\,(\ref{eq:force-y})], we note that
\begin{equation}
S_z(t) \cos \kappa v_{z0} t 
- S_y(t) \sin \kappa v_{z0} t 
= S_z'(t) 
= \frac{ \Omega_1(y(t))
     }{ \tilde \Omega }\ S_{x0}
\,,
\label{eq:dvy-rotating}
\end{equation}
which already evolves slowly. This illustrates that despite the rotating magnetic field, the~SG splitting accumulates over time. The Lorentz force contains oscillating terms [second term in~Eq.(\ref{eq:force-x}, third term in~Eq.(\ref{eq:force-y})] that behave similarly. We exploit the approximation that the transverse velocities~$v_x$, $v_y$ are small compared to~$v_{z0}$ (the ion beam is nearly parallel with the~$z$-axis). Integrating Eq.\,(\ref{eq:force-x}), we find in the leading order a `transverse wiggle' that oscillates at the Doppler frequency:
\begin{equation}
v_x( t ) \approx -\ \frac{ \omega_1( y ) }{ \kappa }\  
                    \cos( \kappa v_{z0} t ) 
   +
   \mbox{constant terms} 
\,.
\label{eq:adiabatic-vx}
\end{equation}
Putting this into the third term of Eq.(\ref{eq:force-y}), we get an expression~$\sim\cos^2(\kappa v_{z0}t)$ with a non-zero time average
(the ponderomotive force). Finally, we find the slow equations of motion for the vertical coordinate
\begin{equation}
\frac{ {\rm d}v_y }{ {\rm d}t } \approx
\omega_0 v_{z}
+ a_{\rm im}( y )
+
\frac{ \omega_1^2( y ) }{ 2 \kappa }
+
u \omega_1( y )\ 
\frac{ \Omega_1( y )
     }{ \tilde \Omega( y ) }\ S_{x0}
\,.
\label{eq:ay-slow}
\end{equation}
Re-instating the acceleration due to the image force, we get Eq.\,(\ref{eq:slow-vertical-force}). To improve the approximation, we combine this equation with~${\rm d}v_z/{\rm d}t \approx-\omega_0 v_y$, keeping only the first term in Eq.\,(\ref{eq:force-z}). With this scheme, we have generated the adiabatic trajectories shown in Fig.\,\ref{fig:bent-paths} (symbols) and Fig.\,\ref{fig:two-blobs}.



\bigskip
\section*{References}

\providecommand{\newblock}{}

\end{document}